\documentclass[
showpacs,preprintnumbers,amsmath,amssymb,floatfix]{revtex4}

\usepackage{graphicx}
\usepackage{dcolumn}
\usepackage{bm}

\usepackage{amsmath}
\usepackage{psfrag}
\usepackage{color}
\usepackage{latexsym,amsfonts}
\providecommand{\href}[2]{#2}   

\def\conj#1{\buildrel{*}\over{#1}}

\def\<{\langle}
\def\>{\rangle}
\def\conj#1{\buildrel{*}\over{#1}}

\begin{document}


\title{Non-nucleon degrees of freedom in the deuteron from the $d(\vec e,e'\vec p\,)n$ break-up}
\author{Alexander~Kobushkin}
\email{kobushkin@bitp.kiev.ua}
\author{Yurij~Kutafin}
\email{kutafin@bitp.kiev.ua}
\affiliation{Bogolyubov Institute for Theoretical Physics\\
Metrologicheskaya str. 14-B, 03143, Kiev, Ukraine}

\date{\today}

\begin{abstract}
We analyzed contribution of quark degrees of freedom in the deuteron to the longitudinal, $P_z'$, and transverse, $P_x'$, polarizations of the proton  in the $d(\vec e,e'\vec p\,)n$ break-up. It is demonstrated that such effects work in correct direction to explain experimental data. We predict that the polarizations should change qualitatively behavior at $p_m \gtrsim$200~MeV/c, (i) the polarizations become strongly dependent on the out-of-plain angle and (ii) there appears a structure in the both polarizations at $p_m \sim$200~MeV/c.
\end{abstract}

\pacs{25.30.Bf, 13.88.+e, 14.20.Gk}
\maketitle

%
Study of the deuteron spin structure at short distances is a hot problem of the modern nuclear physics.
There are strong arguments from experiment~\cite{JINRcs,JINRpol} and theory~\cite{Kobushkin,GlozmanKobSym,Kob,AzhgireyYudin} that exotic non-nucleon degrees of freedom affect strongly the deuteron structure when the internal momentum in the deuteron  $k\gtrsim$200~MeV/c.

For example, one may expect that such effects should be visible in polarization observables in the process of the deuteron electro-disintegration.

Because quarks are fermions one has to take into account the Pauli principal at the level of constituent quarks when the internucleon distance in the deuteron becomes of order of the nucleon size. As a result, at such distances the deuteron wave function includes, apart from the $np$ component, $\mathrm{NN}^\ast$, $\mathrm{N^\ast N}$ and $\mathrm{N^\ast N^\ast}$ components \cite{GlozmanKuchina}. Lowest resonances have negative parity and generate effective $P$-wave in the deuteron which drastically changes behavior of polarization observables of the deuteron break-up at high momentum of a spectator nucleon.

The aim of the present paper is to study how strong new components (generated by the quark degrees of freedom) affect longitudinal, $P_z'$, and  transverse, $P_x'$, polarizations of the proton in the $d(\vec e,e'\vec p\,)n$ reaction at intermediate energy.

\begin{figure}[bt]
\centering
\psfrag{e1}{$e',\, k'$}\psfrag{e}{$e,\, k$}
\psfrag{d}{$d$}
\psfrag{p}{$p$}\psfrag{n}{$n$}\psfrag{NS}{$\mathrm N^\ast$}
\psfrag{+}{$+$}
\includegraphics[width=0.4\textwidth]{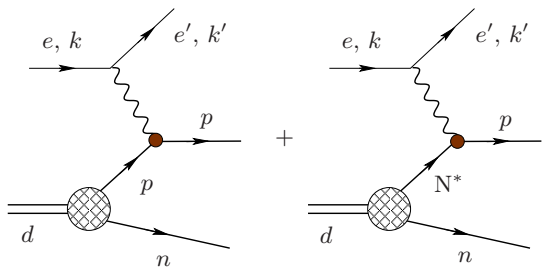}
\caption{\label{fig:mech}Reaction mechanism.}
\end{figure} 

In the framework of the impulse approximation which includes elastic scattering of the electron on the proton and inelastic scattering of the electron on the resonances N$^\ast$ (see Fig.~\ref{fig:mech}) the matrix element reads
\begin{equation}\label{Martix_El}
\begin{split}
&\mathcal M_{sM;s'm m_n} \sim \bar u^{s'}(k')\gamma^\mu u^s(k)\times\\
&\hspace{1.cm}\times\sum_{h\in\{p,\mathrm N^\ast\}}\sum_{m_h}\langle pm|J_\mu|hm_h\rangle \widetilde\Psi_{M m_h m_n}^{(nh)}(\vec n),
\end{split}
\end{equation}
where $u^s(k)$ and $u^{s'}(k')$ are Dirac spinors for the incoming and outgoing electrons with helicity and momentum $s,k$ and $s',k'$, respectively; $\langle pm|J_\mu|hm_h\rangle$ is electromagnetic current matrix element between hadron state ${h}$ and the proton with momenta and spin projections $P', m_h$ and $P, m$, respectively. As the hadron states $h$ we use the proton and the lightest resonances $S_{11}(1535)$ and $D_{13}(1520)$. Obviously, the isospin-$3/2$ resonances do not contribute. In (\ref{Martix_El}) we omitted a common factor, which will be dropped out from final expression for the polarizations. $\widetilde\Psi_{M m_h m_h}^{(nh)}(\vec n)$ is an overlap of the wave function of the deuteron with magnetic quantum number $M$ and the neutron-hadron state $(nh)$. 

Contribution of the electron-neutron scattering was estimated to be negligibly small for discussed further kinematical conditions.

The deuteron wave function, considered at small inter-nuclear distances as a six-quark object, was shown to be qualitatively equivalent to a wave function of the Resonating Group Method (RGM) \cite{GlozmanKuchina,Glozman}
\begin{equation}\label{RGM}
\Psi_M(1,2,\dots,6)=\widehat A\{\varphi_N(1,2,3)\varphi_N(4,5,6)\chi(\vec r\,)\},
\end{equation}
where $\widehat A=\sqrt{\frac1{10}}(1-9\widehat P_{36})$ is a quark antisymmetrizer and $\varphi_N(1,2,3)$ and $\varphi_N(4,5,6)$ are wave functions of nucleon three-quark clusters; $\chi(\vec r\,)$ is the RGM distribution function. As the first step to an appropriate choice of the RGM distribution function we use the conventional NN~deuteron wave function $\psi_M(r)$ and modify it according to the RGM renormalization condition \cite{Oka,Kurdyumov}. For gaussian expansion of the NN~deuteron wave function the overlap reads
\begin{widetext}
\begin{equation}\label{RGM_projection}
\begin{split}
\widetilde\Psi_{M m_h m_n}^{(nh)}(\vec n)=&
\chi_{M m_h m_n}^\text{ren.}(\vec n)\delta_{hp}
+3\sum_{\mu_h}
\langle L_hS_hm_h\mu_h|J_hM_h\rangle\sqrt{\tfrac1{\mathrm{dim}[f_h]}}\gamma_h^X
\sum_k (-1)^{\frac12+S_h+2T_h}A_k I^{L_h}_{N_hL_h00}(n;\alpha_k)\\
&\times Y^\ast_{L_hm_h}(\widehat n)\sqrt{(2T_n+1)(2S_h+1)}
\left\langle S_h \tfrac12\mu_h\mu_n|1M\right\rangle
\sum_{\substack{S_{12}=T_{12}=0,1\\S_{45}=T_{45}=0,1}}
\left\langle [f_h]S_hT_h |[2]S_{12}T_{12};\tfrac12\tfrac12\right\rangle\times\\
&\times
(-1)^{S_{12}+S_{45}}
\left\{
\begin{array}{ccc}
\tfrac12 & T_{12} & \tfrac12\\[0.05cm]
\tfrac12 & T_{45} & \tfrac12
\end{array}
\right\}
\left\{
\begin{array}{ccc}
\tfrac12 & S_{12}  & S_h\\[0.05cm]
S_{45}   &\tfrac12 &\tfrac12 \\[0.05cm]
\tfrac12 &\tfrac12 &1
\end{array}
\right\},
\end{split}
\end{equation}
where $\vec n$ is the internal momentum in the deuteron and $\chi_{M m_h m_n}^\text{ren.}(\vec n)$ is the Fourier transform of the RGM renormalized deuteron wave function
\begin{equation}\label{gaussian}
\begin{split}
&\chi_{M m_h m_n}^\text{ren.}(\vec r\,)=\textstyle{\sqrt{\frac1{4\pi}}}\left\langle\tfrac12\tfrac12m_hm_n|1M\right\rangle u^\text{ren.}(r)+
\sum_{\mu,\xi}\left\langle \tfrac12\tfrac12m_hm_n| 1\mu\right\rangle\langle12\mu\xi|1M\rangle Y_{2\xi}(\widehat r\,)w(r),\\
&u^\text{ren.}(r)=\sum_{k=1}^n A_ke^{-\alpha_k r^2},\quad w(r)=r^2\sum_{k=1}^n B_ke^{-\beta_k r^2}.
\end{split}
\end{equation}
\end{widetext}
For the Paris potential the parameters $A_k$, $\alpha_k$ and $B_k$, $\beta_k$ are given in Ref.~\cite{GlozmanKobSym}. For the further computations with the lightest resonances $S_{11}(1535)$ and $D_{13}(1520)$ we needs only $I^0_{00, 00}$ and $I^1_{11, 00}$, which are given in Ref.~\cite{GlozmanKobSym}
\begin{equation}\label{I000000}
\begin{split}
I^0_{00, 00}(n;\alpha_k)=&\sqrt{2}\left ( \frac{27}{18+30\alpha _k b^2}\right ) ^{3/2} b^3 
\exp \left \lbrace \frac{b^2 n^2}{12} \frac{15 + 16 \alpha _k b^2}{3 + 5 \alpha _k b^2} \right \rbrace ,\\
I^1_{11, 00}(n;\alpha_k) =& I^0_{00, 00}(n;\alpha_k) g_1 bn , \\
g_1 =& \frac{4\alpha _k b^2 - 3}{15 + 16\alpha _k b^2},
\end{split}
\end{equation}
where $b$ is the quark radius of the proton. For the explanation of other notations in (\ref{RGM_projection}) see Ref.~\cite{GlozmanKuchina}.

\begin{figure}[b]
\centering
\psfrag{X}{$x$}
\psfrag{Y}{$y$}
\psfrag{Z}{$z$}
\psfrag{k}{$\vec k$}
\psfrag{k'}{$\vec k'$}
\psfrag{q}{$\vec q$}
\psfrag{p}{$\vec p$}
\psfrag{n}{$\vec n$}
\psfrag{f}{$\phi$}
\includegraphics[width=0.4\textwidth]{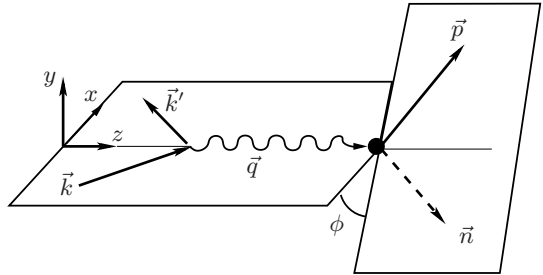}
\caption{\label{fig:kin}The laboratory polarization frame. The $z$ and $y$ axises are chosen along the momentum of virtual photon $\vec q$ and the vector $\vec k \times \vec k'$, respectively. The $x$ axis is in the electron scattering plane completing the right-handed system.}
\end{figure}

The reaction matrix element is defined in the laboratory polarization frame, where the $z$ axis is directed along the momentum transfer $\vec q$, the $y$ axis is in the direction $\vec k \times \vec k'$ and the $x$ axis is in the electron-scattering plane completing the right-hand system, see Fig.~\ref{fig:kin}.

In the hadron rest frame current matrix elements between a hadron state $h$ and the proton are expressed by three independent helicity amplitudes $f_\lambda^{(h)}$ by
\begin{equation}
\begin{split}
\conj{\varepsilon}_\mu\!^{\lambda}\langle p\tfrac12|J^\mu| h\Lambda\rangle &=f_\lambda^{(h)}\delta_{\Lambda,\lambda+\tfrac12},\\
\conj{\varepsilon}_\mu\!^{\lambda}\langle p-\tfrac12|J^\mu| h-\Lambda\rangle &=\eta_hf_\lambda^{(h)}\delta_{\Lambda,\lambda+\tfrac12},
\end{split}
\end{equation}
where $\eta_h=\pi_he^{i\pi\left(s_h-\tfrac12\right)}$, $s_h$ and $\pi_h$ are hadron spin and parity and $\lambda$ is hadron spin projection onto direction of the proton momentum $\vec P$. For resonances with spin $\tfrac12$ the number of independent amplitudes is reduced to two.

Taking into account the orthogonality relations for the polarization vectors $\varepsilon_\mu^{\lambda}$ and the conservation of the electromagnetic current $\langle p\lambda_p|J^\mu| h\Lambda\rangle$ ($\lambda_p$ is the helicity of the proton) one gets the following expressions for the current matrix elements
\begin{equation}
\begin{split}
\langle p\tfrac12|J_\mu| h\Lambda\rangle  & =(-1)^{\Lambda-\tfrac12}\varepsilon^{(\Lambda-1/2)}
_\mu f_{\Lambda-1/2}^{(h)},\\
\langle p-\tfrac12|J_\mu| h\Lambda\rangle& =(-1)^{\Lambda+\tfrac12}\eta_h\varepsilon^{(\Lambda+1/2)}
_\mu f_{-(\Lambda+1/2)}^{(h)}.
\end{split}\label{MElements}
\end{equation}
The polarizations $P_z'$ and $P_x'$ are defined in the polarization frame.
\begin{figure*}[t]
\centering
\includegraphics[width=0.45\textwidth]{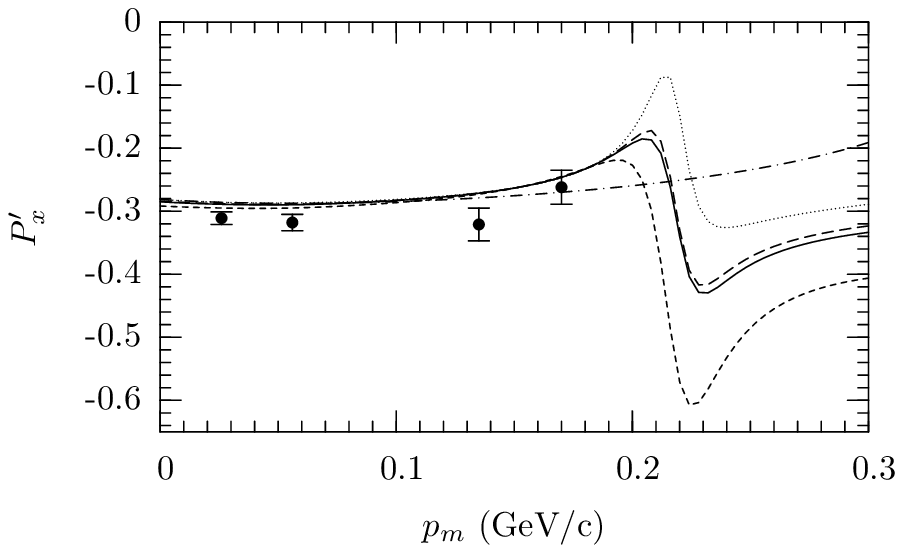}
\includegraphics[width=0.45\textwidth]{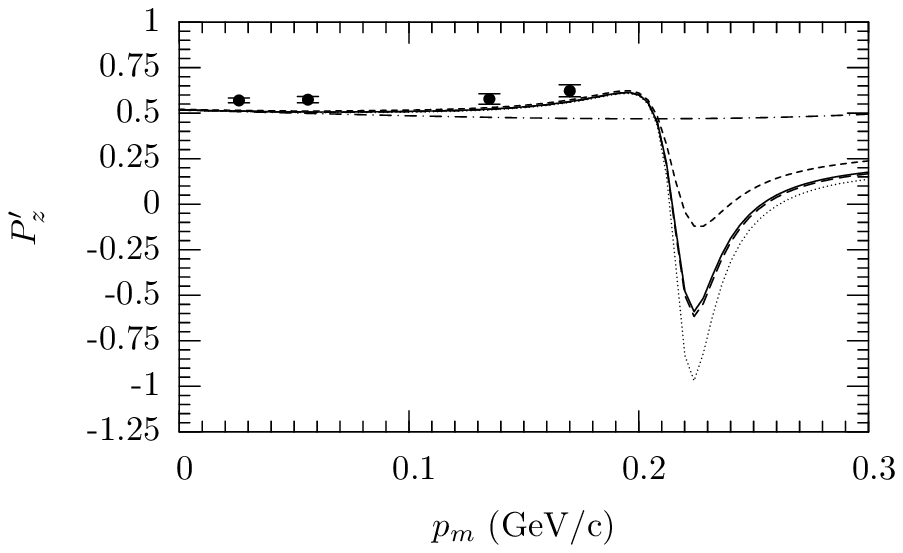}\\
\includegraphics[width=0.45\textwidth]{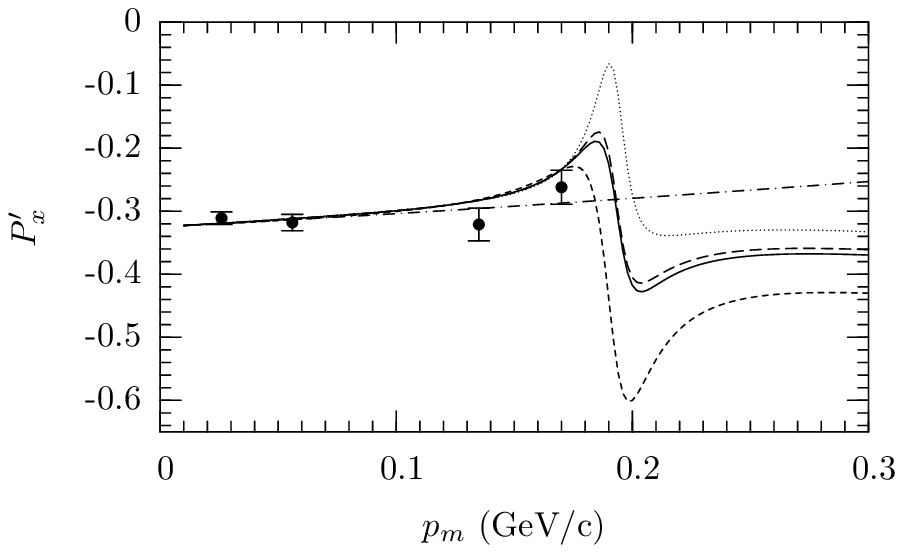}
\includegraphics[width=0.45\textwidth]{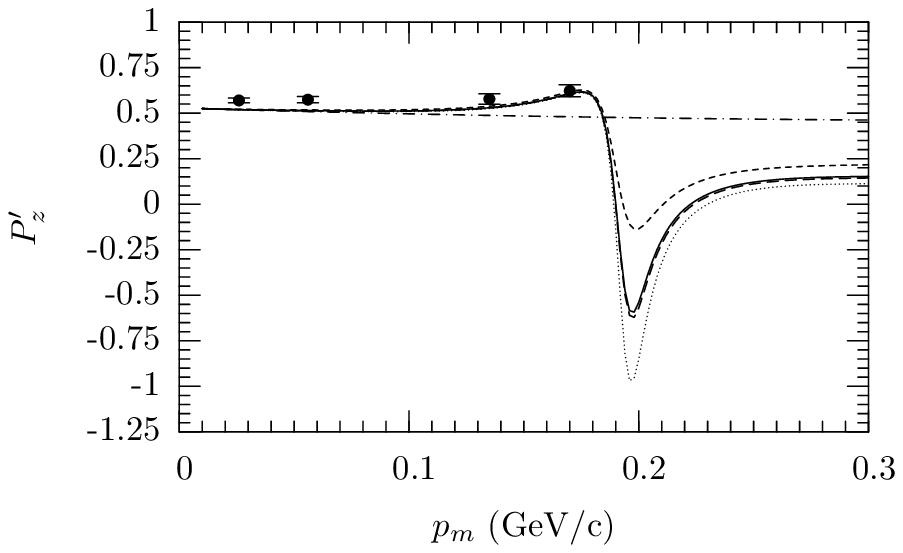}
\caption{Dependence of the transverse, $P_x'$, and longitudinal, $P_z'$, polarizations on the  missing momentum $p_m$. The upper panels are for the deuteron wave function for Paris potential \cite{Paris} and the lower panels are for the NIJM-2 potential \cite{NIJM}. Dash-dotted lines are for the impulse approximation, short dashed, long dashed and dotted lines are for out-of-plain angle $\phi=$0.57$^\circ$, 23.5$^\circ$, 46.4$^\circ$, respectively.  Solid lines are for average polarizations over $\phi$ region between $0^\circ$ to $50^\circ$; data are from Ref.~\cite{Hu_et_al}.}
\label{fig:polar}
\end{figure*}

The expressions (\ref{MElements}) are obviously relativistic covariant and to come to  the polarization frame one has to put
\begin{equation}
\varepsilon^{(0)}_\mu=\frac1{\sqrt{-q^2}}(|\vec q\,|,0,0,q^0),\quad
\varepsilon^{(\pm)}_\mu=\sqrt{\tfrac12}(0,\pm1,-i,0).
\label{pol_vecs}
\end{equation}
The matrix elements between the states with spin projection along the $z$ axis $m_p$ and $m_h$ are defined to be \cite{Berestetski}
\begin{equation}
\begin{split}
&\langle p m_p|J_\mu|hm_h\rangle=\\
&=\sum_{\lambda_p,\Lambda}\mathcal D_{\lambda_p m_p}^{(1/2)}(\varphi,\theta,0)\conj{\mathcal D}_{\Lambda m_h}\hspace{-0.55cm}^{(j_h)}(\varphi,\theta,0)\langle p\lambda_p|J_\mu| h\Lambda\rangle,
\end{split}\label{Wigner}
\end{equation}
where $\varphi$ and $\theta$ are polar and azimuthal angles of the proton momentum in the hadron frame and $\mathcal D_{\Lambda m_h}^{(j_h)}(\varphi,\theta,0)$ is the Wigner D-function.

The polarization transfer along the direction $\vec \xi$ is defined to be
\begin{equation}
P'_{\vec \xi}=\frac{\sigma^+-\sigma^-}{\sigma^++\sigma^-}\;,
\label{definition}
\end{equation}
where $\sigma^{\pm}$ is the differential cross section for the reaction $d(\vec e, e'\vec p\,)n$ with right-hand polarized electron and the proton polarized along (opposite to) $\vec \xi$. To calculate this cross sections one has to come from the amplitudes (\ref{Martix_El}) for the proton polarized along the $z$ axis to amplitudes for the proton polarized along the axis $\vec \xi=(\sin\theta '\cos\varphi',\sin\theta'\sin\varphi',\cos\theta')$ parametrized by two angles $\theta'$ and $\varphi'$ 
\begin{equation}
\begin{split}
&\mathcal M_{sM;s'+\frac12m_n}^{(\vec \xi\,)}
=\mathcal M_{sM;s'+\frac12m_n}e^{\frac{i}2\varphi '}\cos\tfrac{\theta '}{2}+ \mathcal M_{sM;s'-\frac12m_n}e^{-\frac{i}2\varphi '}\sin\tfrac{\theta '}{2},\\
&\mathcal M_{sM;s'-\frac12m_n}^{(\vec \xi\,)}
=-\mathcal M_{sM;s'+\frac12m_n}e^{\frac{i}2\varphi '}\sin\tfrac{\theta '}{2} +\mathcal M_{sM;s'-\frac12m_n}e^{i\frac{i}2\varphi '}\cos\tfrac{\theta '}{2}
\end{split}
\label{definition1}
\end{equation}
and
\begin{equation}
\sigma^{\pm} \sim \sum_{Ms'm_n}\left|\mathcal M_{\frac12M;s'\pm\frac12m_n}^{(\vec \xi\,)}\right|^2.
\label{definition2}
\end{equation}
For the longitudinal and transverse polarizations the vector $\vec \xi$ is specified as
\begin{equation*}
\begin{array}{ll}
\theta '=0, \varphi '\ \text{is arbitrary} \quad &\text{--- for longitudinal polarization},\\
\theta '=\tfrac{\pi}2,\ \varphi '=0 \quad &\text{--- for transverse polarization}.
\end{array}
\end{equation*}

There are five independent kinematical variables, the energy of initial and final electron, $E$ and $E'$, the electron scattering angle, $\theta_e$, the neutron momentum $n$ and the ``out-of-plane'' angle  $\phi$ (the angle between the electron-scattering plane and the proton-neutron plane, see Fig.~\ref{fig:kin}). In Figs.~\ref{fig:polar} the results for  the $P_z'$ and $P_x'$ polarizations are compared with experimental data of Ref.~\cite{Hu_et_al} measured at $E=1.669$~GeV, $E'=1.127$~GeV, the electron scattering angle, $\theta_e=42.65^\circ$. The ``missing momentum'' $p_m$ of Ref.~\cite{Hu_et_al} is identified with the neutron momentum $n$. Different curves in Figs.~\ref{fig:polar} correspond to different angles $\phi$. In numerical calculations the helicity amplitudes $f_\lambda^{(p)}$ were expressed in terms of the proton electric and magnetic form factors and for the resonance amplitudes we used the parametrization \cite{BK}.

In our calculations we did not take into account MC and IC effects, as well as final-state interaction between outgoing particles. Nevertheless, one may conclude the following:
\begin{itemize}
\item Effects connected with quark structure of the deuteron work in correct direction to explain data and should change qualitatively behavior of the polarized observables at $p_m \gtrsim$200~MeV/c.
\item Starting from $p_m\sim$200~MeV/c the polarizations become strongly dependent on the angle $\phi$.
\item Position of a structure in the polarization observables are slightly dependent on the choice of two nucleon potential. For Paris and NIJM2 potentials the structure appears at $p_m\approx$230 and 200~ MeV/c, respectively.
\end{itemize}

This features may be a strong test for experimental observation of the deuteron quark structure at short distances.

\begin{acknowledgements}
The authors grateful for Charles Perdrisat for discussion of the results of experiment \cite{Hu_et_al} and Eugene Strokovsky for reading manuscript.
\end{acknowledgements}
%

\end{document}